\documentclass[12pt]{article}
\usepackage{epsfig}
\usepackage{bm}

\begin{document}

\title{Prime numbers and spontaneous neuron activity}

\author{\small A. Bershadskii\\
\small ICAR, P.O. Box 31155, Jerusalem 91000, Israel\\}

\date{}
\maketitle

\begin{abstract}
Logarithmic gaps have been used in order to find a periodic component of the sequence 
of prime numbers, hidden by a random noise (stochastic or chaotic). It is shown that 
multiplicative nature of the noise is the main reason for the successful application of the 
logarithmic gaps transforming the multiplicative noise into an additive one. A relation of this 
phenomenon to spontaneous neuron activity and to chaotic brain computations has been discussed.

\end{abstract}

\maketitle

\section{Introduction}

The prime number distribution is apparently random. The apparent randomness 
can be stochastic or chaotic (deterministic). It is well known that the Riemann zeros obey the
chaotic GUE statistics (if the primes are interpreted as the classical periodic orbits of a 
chaotic system, see for a recent review \cite{bogom}), whereas the primes themselves 
are believed to be stochastically distributed (Poissonian-like etc. \cite{sou}). 
However, recent investigations suggest that primes themselves 
"...could be eigenvalues of a quantum system whose classical counterpart is chaotic at low energies 
but increasingly regular at higher energies." \cite{tt}. 
Therefore, the problem of chaotic (deterministic)
behavior of the moderate and small prime numbers is still open. Moreover, there are also 
an indication of periodic patterns in the prime numbers distribution \cite{tt}-\cite{ac}. These patterns, 
however, has been observed in {\it probability} distribution of the gaps between neighboring primes and 
not in the prime numbers sequence itself (see also Ref. \cite{dah}). The intrinsic randomness 
(stochastic or chaotic) of the primes distribution 
makes the problem of finding the periodic patterns in the prime numbers themselves 
(if they exist after all) a very difficult one. 

A physicists may ask: Why should one be interested in finding these patterns? The answer is: comparison. 
If one can recognize patterns in an apparently random system, then one can compare these patterns with 
the patterns known for some other systems of interest. We have already mentioned the comparison with certain 
quantum systems. Another intrinsic comparison can be made with the computational properties of brains, 
where the natural numbers certainly should play a crucial role. The neuron signals are also apparently 
random. Can one compare patterns observed in these signals with the patterns in the
prime numbers sequence in a constructive way, in order to shed a light on the computational apparatus of the 
brains?

\section{Logarithmic gaps}

It is believed that properties of the gaps between consecutive primes can provide a lot of information about 
the primes distribution in the natural sequence. The so-called prime number theorem states that 
the "average length" of the gap between a prime $p$ and the next prime number is proportional (asymptotically) 
to $\ln p $ (see, for instance, Ref. \cite{sou}). This implies, in particular, statistical non-stationarity of the prime numbers sequence. That is 
a serious obstacle for practical applications of the statistical methods to this sequence. For relatively 
large prime numbers one can try to overcome this obstacle by using relatively short intervals \cite{sou}. 
Another (additional) way to overcome this obstacle is to use logarithmic gaps (logarithms of the gaps). 
The non-stationarity in the sequence of the logarithmic gaps is considerably 'slower' than that in the original sequence of the gaps themselves. The logarithmic gaps have also another crucial advantage. If the noise 
corrupting the gaps sequence has a multiplicative nature, then for the logarithmic gaps this noise 
will be transformed into an additive one. It is well known that the additive noises can be much readily 
separated from the signal then the multiplicative ones (see below).

 Before starting the analysis let us recall two rather trivial properties of the gaps, which will be used below. 
The sequence can be restored from the gaps by taking cumulative sum. Pure periodicity in a sequence 
corresponds to a constant value of the gaps (the period). 

Let us take, as a first step, logarithms of the gaps for the sequence of prime numbers. Then, let us compute 
cumulative sum of these logarithms as a second step. 
Then, we will multiply each value in the cumulative sum by 10 and will replace each of the obtained values 
by a natural number which is nearest to it. As a result we will obtain a sequence of natural numbers: 
7,14,28,35,49,55,.... which will be called as ln-sequence.

Let us now define a binary function $v(n)$ of natural numbers $n=2,3,4,...$, which 
takes two values +1 or -1 and changes its sign passing any number from the ln-sequence. This function contains 
full information about the ln-sequence numbers distribution in the sequence of natural numbers.\\ 

\begin{figure} \vspace{-2cm}\centering
\epsfig{width=.7\textwidth,file=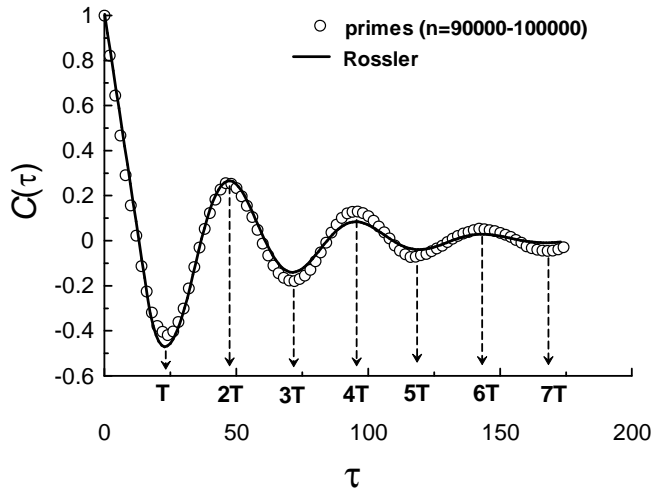} \vspace{-6.5cm}
\caption{Autocorrelation of the $v(n)$ functions for the ln-sequence (circles) in the interval $9\times 10^4 < n < 10^5$. The solid line correspond to autocorrelation function computed for the telegraph signal 
generated by the R\"{o}ssler attractor fluctuations overcoming the threshold $x = 7$. In order to make the autocorrelation functions comparable a rescaling has been made for the R\"{o}ssler autocorrelation function. }
\end{figure}
 
 Let us consider a relatively short interval: $9\times 10^4 < n < 10^5$. For this interval the autocorrelation 
function 
$$
C(n, \tau)= \langle v(n)v(n + \tau)\rangle - \langle v(n)\rangle \langle v(n + \tau)\rangle  \eqno{(2.1)}
$$
computed for the $v(n)$ function of the ln-sequence will be approximately independent on $n$. Figure 1 
shows the autocorrelation function (circles) computed for the $v(n)$ function of 
the ln-sequence for the 'short interval': $9\times 10^4 < n < 10^5$. Figure 2 shows an  autocorrelation function (circles) computed for the $v(n)$ function of the ln-sequence for interval of much smaller natural numbers: $1000 < n < 2\times 10^4$ (here we rely on the much slower 
non-stationarity in the logarithmic gaps sequence in comparison with the original gaps sequence). 
\begin{figure} \vspace{-2cm}\centering
\epsfig{width=.7\textwidth,file=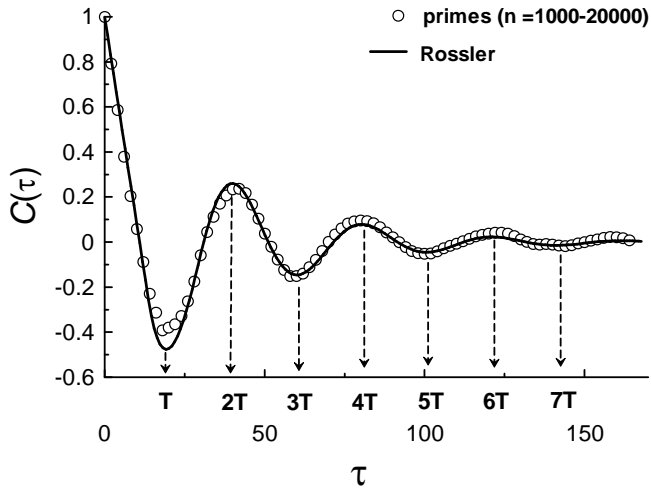} \vspace{-6.5cm}
\caption{As in Fig. 1 but for interval $1000 < n < 2\times 10^4$.
 }
\end{figure}

\section{Models}
\begin{figure} \vspace{-2cm}\centering
\epsfig{width=.7\textwidth,file=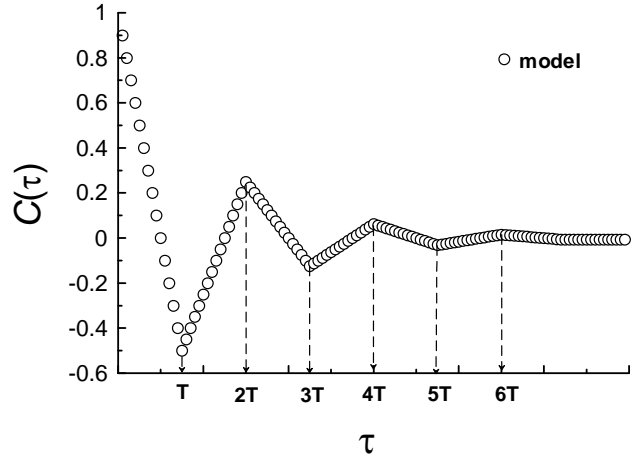} \vspace{-6.5cm}
\caption{Autocorrelation function for the simple model telegraph signal: Eq. (3.2) with $q=0.25$. 
 }
\end{figure}
In order to understand origin of the oscillating autocorrelation function shown in Figs. 1 and 2
let us consider a very simple telegraph signal, which allows analytic calculation of its 
autocorrelation function. The telegraph signal takes two values +1 or -1 and it changes its sign at 
discrete moments: 
$$
t_n=nT+\zeta \eqno{(3.1)}
$$
where $\zeta$ is an uniformly distributed over the interval $[0, T]$ random variable, 
$n=1,2,3...$ and $T$ is a fixed period. If $q$ 
is a probability of a sign change at a current moment ($0 \leq q < 1$), then the autocorrelation 
function of such telegraph signal is:
$$
C(\tau) = (n-\tau/T)(2q-1)^{n-1}+(\tau/T-(n-1))(2q-1)^n  \eqno{(3.2)}
$$
in the interval $(n-1)T \leq \tau < nT$. Figure 3 shows the autocorrelation function Eq. (3.2) calculated 
for $q=0.25$, as an example (cf. with Figs. 1 and 2).
\begin{figure} \vspace{-2cm}\centering
\epsfig{width=.7\textwidth,file=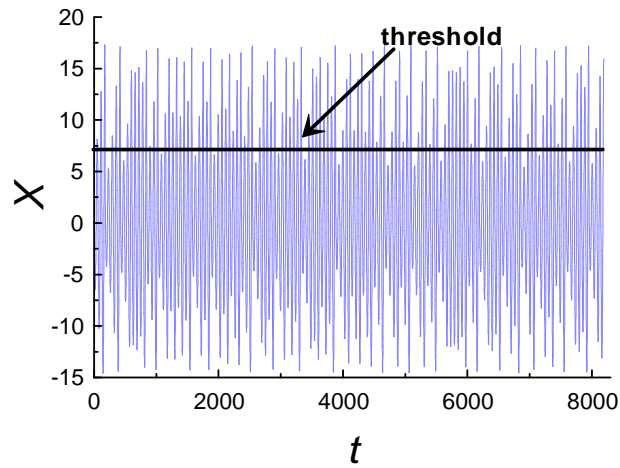} \vspace{-6.5cm}
\caption{X-component fluctuations 
of a chaotic solution of the R\"{o}ssler system Eq. (3.3) ($a=0.15,~ b=0.20,~ c=10.0$). }
\end{figure}
In the model Eq. (3.1) $\zeta$ was taken as a pure stochastic variable. This variable, however, can be also a chaotic (deterministic) one. Let us consider a chaotic solution of the R\"{o}ssler system \cite{ros} 
$$
\frac{dx}{dt} = -(y + z);~~  \frac{dy}{dt} = x + a y;~~  \frac{dz}{dt} = b + x z - c z\eqno{(3.3)}
$$      
where a, b and c are parameters (a reason for this system relevance will be given in Section 4). 
Figure 4 shows the x-component fluctuations of a chaotic solution 
of the R\"{o}ssler system.  

Let us consider a telegraph signal generated by the variable $x$ crossing certain threshold from below (Fig. 4). 
This telegraph signal takes values: +1 or -1, and changes its sign at the threshold-crossing points. 
Then, the fundamental period of the R\"{o}ssler chaotic attractor provides the period $T$ in Eq. (3.1), 
whereas the chaotic fluctuations of the variable $x$ of the attractor provide chaotic variable 
$\zeta$ in the Eq. (3.1) for the considered telegraph signal. Autocorrelation function computed for this telegraph 
signal has been shown in Figs. 1 and 2 as the solid line. In order to make the autocorrelation functions comparable a rescaling has been made for the R\"{o}ssler telegraph signal's autocorrelation function: the
scaling coefficient is equal to 0.36 for Fig. 1 and to 0.30 for Fig. 2.

\section{Prime numbers and spontaneous neuron activity}

\begin{figure} \vspace{-2cm}\centering
\epsfig{width=.6\textwidth,file=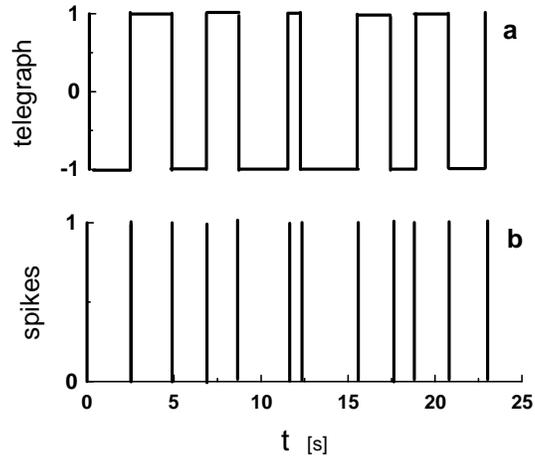} \vspace{-5cm}
\caption{Mapping of a spike train (figure 1b) into a telegraph signal (figure 1a).  }
\end{figure}

All types of information, which is received 
by sensory system, are encoded by nerve cells 
into sequences of pulses of similar shape 
(spikes) before they are transmitted to the 
brain. Brain neurons use such 
sequences as main instrument for intercells 
connection. The information is reflected in 
the time intervals between successive firings 
(interspike intervals of the action potential 
train, see Fig. 5b). There need be no loss of 
information in principle when converting from 
dynamical amplitude information to spike trains 
and the irregular spike sequences 
are the foundation of neural information 
processing. Although understanding of the origin 
of interspike intervals irregularity has important 
implications for elucidating the temporal 
components of the neuronal code the problem is still very far from 
its solution.

\begin{figure} \vspace{-2cm}\centering
\epsfig{width=.6\textwidth,file=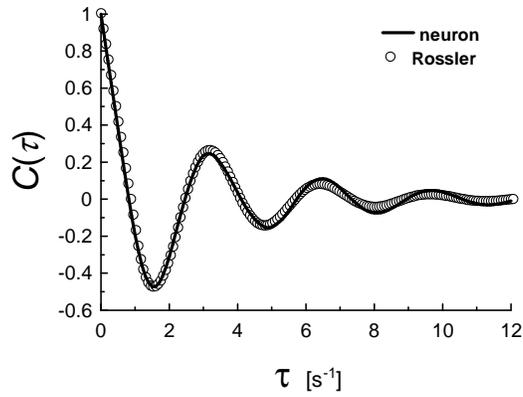} \vspace{-6cm}
\caption{Autocorrelation functions for the telegraph signals corresponding to a spontaneous activity of 
a hippocampal (CA3) singular neuron (solid curve) and to the spike train generated by the R\"{o}ssler attractor fluctuations overcoming a threshold (circles). 
In order to make the autocorrelation functions comparable a rescaling has been made 
for the R\"{o}ssler autocorrelation function. The data has been taken from the Ref. \cite{bi}. 
 }
\end{figure}
In a recent paper \cite{bi} a membrane threshold-crossing model of spontaneous (without external stimulus) neuron firing based on the chaotic R\"{o}ssler attractor was suggested. Good agreement between the model and the spiking time-series, obtained in vitro from a spontaneous activity of hippocampal (CA3) singular neurons (rat's brain slice culture), was reported in this paper. 
The spiking time-series were mapped into a telegraph signal as it is shown in Fig. 5. 
The comparison between the experimental data and the signal generated by the R\"{o}ssler attractor fluctuations overcoming a threshold (cf. previous Section) is shown in Fig. 6.  
\begin{figure} \vspace{-2cm}\centering
\epsfig{width=.7\textwidth,file=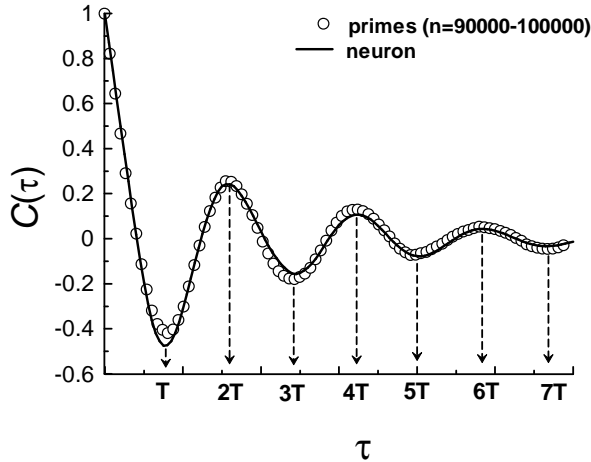} \vspace{-7cm}
\caption{Autocorrelation of the $v(n)$ functions for the ln-sequence (circles) in the interval $9\times 10^4 < n < 10^5$. The solid curve corresponds to autocorrelation function computed for the telegraph signals corresponding to a spontaneous activity of a hippocampal (CA3) singular neuron (cf. Fig. 3).
In order to make the autocorrelation functions comparable a rescaling has been made for autocorrelation of the $v(n)$ function. }
\end{figure}
\begin{figure} \vspace{-2cm}\centering
\epsfig{width=.7\textwidth,file=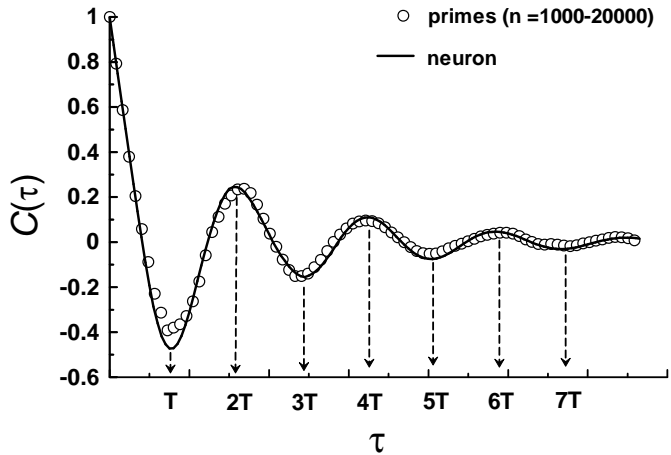} \vspace{-7cm}
\caption{As in Fig. 7 but for interval $1000 < n < 2\times 10^4$.  }
\end{figure}
  
The hippocampus is a significant part of a brain system responsible, in particular, for spatial memory, body orientation and navigation. These functions of brain are related to the very sophisticated real-time computations, which  the brains make with astonishing effectiveness. One should distinguish between the inner brain computations 
and our conscious arithmetic calculations, which are performed in the more specialized brain's departments such as the intraparietal sulcus \cite{deh}. Therefore, if one wants to investigate how the inner brain computational abilities are related to neuron activity the hippocampal pyramidal cells are an appropriate subject for this investigation.  Spontaneous activity in brain slice preparations purely reflects the intrinsic properties of local circuits 
and individual neurons. The comparison of the spontaneous activity of the hippocampal singular neurons with the random properties of the sequence of prime numbers can shed certain light on the long standing problem: whether mathematics belongs completely to the outer world or it is a production of our brain activity 
(at least in the form of the natural numbers). For instance, in order to work together the brain neurons 
have to make an adjustment of their rhythms. The main problem for this adjustment is the very noisy environment 
of the brain neurons \cite{all}. For pure periodic inner clocks this adjustment would be impossible due to the noise. 
Nature, however, has another option. This option is a chaotic clock (see Ref. \cite{bi} and references therein). In chaotic attractors certain characteristic frequencies can be embedded by broad-band spectra, that makes them much more stable to the noise perturbations. The Riemann's hypothesis, for instance, which states that the apparent randomness in the distribution of prime numbers only comes from a noise (otherwise the prime numbers are as regularly distributed as possible), can reflect an optimization of such mechanism. 

On the other hand, the compatibility of the statistical properties of the neuron spontaneous activity  with the 
statistics of the prime numbers (see Figs. 7 and 8) could be a key for understanding of the inner computational 
methods of the brains. If brains use chaotic dynamics (Fig. 6) also for 
computational processing \cite{lb}, then such compatibility should play a crucial role in the inner brain computations. 
The reduction to the additive random processes by means of the logarithmic gaps (see previous Section) could be a part of these inner brain computations. If the gaps can be considered as a result of a multiplicative random process, i.e. each gap between consecutive primes can be represented as a product of independent random variables (the random variables could take their values from the prime numbers set, including 1), then logarithmic gap represents the sum of random variables (i.e. an additive random process). It is known that unlike the additive random processes the multiplicative random processes are much more complex and difficult for analysis.  \\

I thank Y. Ikegaya for sharing the neuron data and discussions.


\newpage
\begin{thebibliography}{99}
\bibitem{bogom} E. Bogomolny, Riemann zeta function and quantum chaos, 
Prog. Theor. Phys. Supplement, vol. 166,  pp. 19-44, 2007.
\bibitem{sou} K. Soundararajan, The distribution of prime numbers, NATO Science Series II: 
Mathematics, Physics and Chemistry, , vol. 237, pp. 59-83, 2007.
\bibitem{tt} T. Timberlake and J.Tucker, Is there quantum chaos in the prime numbers?, 
Bulletin of the American Physical Society,  vol. 52, p. 35, 2007, (arXiv:0708.2567).
\bibitem{po} C.E. Porter, Statistical Theories of Spectra: Fluctuations, NY: Academic Press, 1965.
\bibitem{wo} M. Wolf, Applications of statistical mechanics in
number theory, Physica A, vol. 274, pp. 149-157, 1999.
\bibitem{kis} P. Kumar, P.Ch. Ivanov, and H.E. Stanley, Information Entropy and Correlations in Prime Numbers, 
arXiv:cond-mat/0303110.
\bibitem{ac} S. Ares, and M. Castro, Hidden structure in the randomness
of the prime number sequence?, Physica A, vol. 360, pp. 285-296, 2006.
\bibitem{dah} S. R. Dahmen, S. D. Prado and T. Stuermer-Daitx, Similarity in the statistics of prime numbers, 
Physica A, vol. 296, pp. 523-528, 2001. 
\bibitem{ros} O.E. R\"{o}ssler, An equation for continuous chaos, Phys. Lett. A,  vol. 57, pp. 397-398, 1976.
\bibitem{bi} A. Bershadskii, Y. Ikegaya, Chaotic neuron clock, Chaos, Solitons \& Fractals, vol. 44, pp. 
342-347, 2011.
\bibitem{deh} S. Dehaene, Arithmetic and the brain, Current Opinion in Neurobiology, vol. 14, 
pp. 218-224, 2004. 
\bibitem{all} P. Allegrini, D. Menicucci, R. Bedini, A. Gemignani, and P. Paradisi, Complex 
intermittency blurred by noise: theory and application to neural dynamics,
Phys. Rev. E, vol. 82, 015103(R) 2010.
\bibitem{lb} R. Legenstein and W. Maass, What makes a dynamical system computationally powerful? in 
"New Directions in Statistical Signal Processing: From Systems to Brain", S. Haykin et. al. editors,  
pp. 127-154 (MIT Press, Cambridge, MA, 2007).

\end{thebibliography}
\end{document}